\documentclass[aps,prl,twocolumn,superscriptaddress]{revtex4}
\usepackage{eurosym}
\usepackage{amsfonts}
\usepackage{amssymb}
\usepackage{amsmath}
\usepackage{graphicx}
\usepackage{epstopdf}
\usepackage{color}
\usepackage[colorlinks=true,urlcolor=blue,linkcolor=blue,citecolor=blue]{hyperref}

\begin{document}

\title{Temperature-induced Lifshitz transition and possible excitonic
instability in ZrSiSe}
\author{F. C. Chen}
\thanks{These authors have contributed equally to this work.}
\affiliation{Key Laboratory of Materials Physics, Institute of Solid State Physics,
HFIPS, Chinese Academy of Sciences, Hefei, 230031, China}
\affiliation{University of Science and Technology of China, Hefei, 230026, China}
\author{Y. Fei }
\thanks{These authors have contributed equally to this work.}
\affiliation{Department of Physics, Zhejiang University, Hangzhou, 310027, China}
\author{S. J. Li}
\thanks{These authors have contributed equally to this work.}
\affiliation{Institute of Applied Physics and Computational Mathematics, Beijing, 100088,
China}
\affiliation{Anhui Province Key Laboratory of Condensed Matter Physics at Extreme
Conditions, High Magnetic Field Laboratory, HFIPS, Chinese Academy of Sciences,
Hefei, 230031, China}
\affiliation{College of Mathematics and Physics, Beijing University of Chemical Technology, Beijing 100029, China}
\author{Q. Wang}
\affiliation{Anhui Province Key Laboratory of Condensed Matter Physics at Extreme
Conditions, High Magnetic Field Laboratory, HFIPS, Chinese Academy of Sciences,
Hefei, 230031, China}
\affiliation{University of Science and Technology of China, Hefei, 230026, China}
\author{X. Luo }
\email{xluo@issp.ac.cn}
\affiliation{Key Laboratory of Materials Physics, Institute of Solid State Physics,
HFIPS, Chinese Academy of Sciences, Hefei, 230031, China}
\author{J. Yan}
\affiliation{Key Laboratory of Materials Physics, Institute of Solid State Physics,
HFIPS, Chinese Academy of Sciences, Hefei, 230031, China}
\affiliation{University of Science and Technology of China, Hefei, 230026, China}
\author{W. J. Lu}
\affiliation{Key Laboratory of Materials Physics, Institute of Solid State Physics,
HFIPS, Chinese Academy of Sciences, Hefei, 230031, China}
\author{P. Tong}
\affiliation{Key Laboratory of Materials Physics, Institute of Solid State Physics,
HFIPS, Chinese Academy of Sciences, Hefei, 230031, China}
\author{W. H. Song}
\affiliation{Key Laboratory of Materials Physics, Institute of Solid State Physics,
HFIPS, Chinese Academy of Sciences, Hefei, 230031, China}
\author{X. B. Zhu}
\affiliation{Key Laboratory of Materials Physics, Institute of Solid State Physics,
HFIPS, Chinese Academy of Sciences, Hefei, 230031, China}
\author{L. Zhang}
\affiliation{Anhui Province Key Laboratory of Condensed Matter Physics at Extreme
Conditions, High Magnetic Field Laboratory, HFIPS, Chinese Academy of Sciences,
Hefei, 230031, China}
\author{H. B. Zhou}
\affiliation{Anhui Province Key Laboratory of Condensed Matter Physics at Extreme
Conditions, High Magnetic Field Laboratory, HFIPS, Chinese Academy of Sciences,
Hefei, 230031, China}
\author{F. W. Zheng}
\affiliation{Institute of Applied Physics and Computational Mathematics, Beijing, 100088,
China}
\author{P. Zhang}
\affiliation{Institute of Applied Physics and Computational Mathematics, Beijing, 100088,
China}
\affiliation{School of Physics and Physical Engineering, Qufu Normal University, Qufu 273165, China}
\author{A. L. Lichtenstein}
\affiliation{Institute for Theoretical Physics, University Hamburg, Jungiusstrasse 9,
D-20355 Hamburg, Germany}
\affiliation{Theoretical Physics and Applied Mathematics Department, Ural Federal
University, Mira Street 19, 620002 Ekaterinburg, Russia}
\author{M. I. Katsnelson}
\affiliation{Theoretical Physics and Applied Mathematics Department, Ural Federal
University, Mira Street 19, 620002 Ekaterinburg, Russia}
\affiliation{Institute for Molecules and Materials, Radboud University, Heijendaalseweg
135, NL-6525AJ Nijmegen, The Netherlands}
\author{Y. Yin }
\email{yiyin@zju.edu.cn}
\affiliation{Department of Physics, Zhejiang University, Hangzhou, 310027, China}
\affiliation{Collaborative Innovation Center of Microstructures, Nanjing University,
Nanjing, 210093, China}
\author{Ning Hao }
\email{haon@hmfl.ac.cn}
\affiliation{Anhui Province Key Laboratory of Condensed Matter Physics at Extreme
Conditions, High Magnetic Field Laboratory, HFIPS, Chinese Academy of Sciences,
Hefei, 230031, China}
\author{Y. P. Sun }
\email{ypsun@issp.ac.cn}
\affiliation{Anhui Province Key Laboratory of Condensed Matter Physics at Extreme
Conditions, High Magnetic Field Laboratory, HFIPS, Chinese Academy of Sciences,
Hefei, 230031, China}
\affiliation{Key Laboratory of Materials Physics, Institute of Solid State Physics,
HFIPS, Chinese Academy of Sciences, Hefei, 230031, China}
\affiliation{Collaborative Innovation Center of Microstructures, Nanjing University,
Nanjing, 210093, China}

\begin{abstract}
The nodal-line semimetals have attracted immense interest due to the unique
electronic structures such as the linear dispersion and the vanishing
density of states as the Fermi energy approaching the nodes. Here, we report
temperature-dependent transport and scanning tunneling
microscopy(spectroscopy)[STM(S)] measurements on nodal-line semimetal ZrSiSe. Our
experimental results and theoretical analyses consistently demonstrate that
the temperature induces Lifshitz transitions at 80 K and 106 K in ZrSiSe,
which results in the transport anomalies at the same temperatures. More
strikingly, we observe a V-shape dip structure around Fermi energy from the
STS spectrum at low temperature, which can be attributed to co-effect of the spin-orbit coupling and excitonic
instability. Our observations indicate the correlation interaction may play
an important role in ZrSiSe, which owns the quasi-two-dimensional electronic
structures.
\end{abstract}

\maketitle

Lifshitz transition, which is characterized by the change of the Fermi
surface topology\cite{1}, can happen among various materials through
changing some external parameters such as chemical doping, pressure or
magnetic field\cite{2, 3, 4, 5, 6, 7}. Recently, a new type of
temperature-driven Lifshitz transition has been observed in Dirac and Weyl
semimetals, in which the transition usually happens as the Fermi energy
crosses the Dirac and Weyl nodes\cite{8, 9, 10}. Near these nodes, the
carriers have high mobility and can switch from $n$($p$)-type to $p$($n$%
)-type as the Fermi energy decreases (increases) to cross the nodes\cite{10}%
. As a consequence, a Lifshitz transition usually has a close relationship
with the transport anomalies at low temperature in Dirac and Weyl semimetals%
\cite{11, 12, 13, 14}. However, no such transition has been observed in
nodal-line semimetal. In terms of materials, the experimentally confirmed
compounds include PbTaSe$_{2}$, PtSn$_{4}$ and ZrSiM (M=S, Se Te) family\cite%
{15, 16, 17, 18, 19, 20, 21}. Among them, ZrSiM attracts much attention due
to the quite large energy window of the linear band dispersion in some
region of the Brillouin zone. As the element M changes from S to Te, the
electronic structures between ZrSiM show some subtle but important
differences. In ZrSiSe and ZrSiTe, some new tiny trivial bands could cross
the Fermi energy. This feature implies the electronic structures of these
two compounds are in the vicinity of a Lifshitz transition, which could be
achieved by tuning some external parameters, such as temperature. Furthermore,
the first-principles calculations predict a spin-orbit coupling (SOC) gap (20$\sim $%
60 meV), which can destroy the nodal-line structure\cite{15, 18, 21}. When
the correlation interaction is taken into account, the nodal-line structure
can further become instable and various correlation-driven nodal-line
instabilities could emerge such as mass enhanced effect, excitonic
insulator, charge/spin density wave etc\cite{22, 23, 24}. These destructive
instabilities could break down and obscure the topological physics based on
the nodal-line structures in ZrSiM. However, the experimentally testing and
verifying the destructive instabilities is still lacking.

In this work, we synthesize the high-quality single crystal ZrSiSe, and
perform the detailed transport and scanning tunneling
microscopy(spectroscopy) [STM(S)] measurements. The technical details are
present in Supplemental Materials (SMs)\cite{51}. The transport measurements show
that ZrSiSe is a $p$-type metal above a critical temperature $T$= 106 K,
whereas a portion of $n$-type carriers suddenly arise accompanying with the
mobility sharply dropping when the temperature decreases below 106 K. As
temperature further decreases, the mobility of $n$-type carriers suddenly
ascends at another critical temperature $T$=80 K. In combined with the
first-principles calculations, these transport anomalies can be interpreted
by the change of the topology of the Fermi surface induced by Fermi level
shift as temperature changes. Furthermore, our STM/S measurements
demonstrate the carrier density ratio between $n$-type and $p$-type develops
from 0.7 at $T$=77 K to 0.95 at $T$=4.5 K. The charge neutrality from
imbalance to balance as temperature decreases further proves the change of
the topology of the Fermi surface. These self-consistent interpretations to
the independent transport and STM/S measurements strongly indicate the
temperature-dependent Lifshitz transition in ZrSiSe. Through investigating
the average differential conductance spectra from STS measurements, we
observe a V-shape dip structure around Fermi energy. Our theoretical model
and analyses indicate that the dip formation in ZrSiSe is referable to the
co-effect of SOC and the destructive correlation effect, which gives rise to
the excitonic instability. Thus, ZrSiSe is a promising candidate for
studying the destructive effect from correlation effect and SOC to Dirac
nodal-line physics.

\begin{figure}[tbp]
\begin{flushleft}
\includegraphics[width=1\columnwidth]{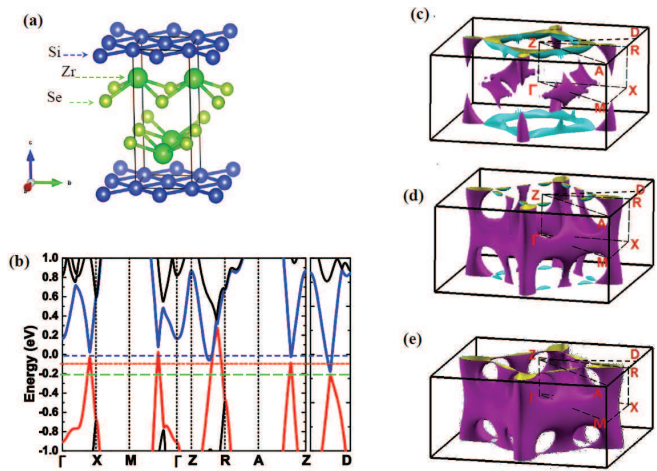}
\caption{(a) Crystal structure of ZrSiSe. (b) The first-principles calculated band structure of ZrSiSe with SOC along high-symmetry lines. The dashed blue line, dotted red and long dashed green lines label three different Fermi levels corresponding to the cases with temperature $T$=5 K (intrinsic case), 102 K and 106 K according to Fig. 2 (e), respectively. (c), (d) and (e) The three-dimensional Fermi surface with the Fermi energy labeled by the dashed blue, dotted red and long dashed green lines in (b), respectively.}
\end{flushleft}
\end{figure}

The crystal structure of ZrSiSe is shown in Fig. 1(a). ZrSiSe has a
PbFCl-type crystal structure with space group $P$4/$nmm$. The band structure
along some momentum lines are plotted in Fig. 1(b). When the Fermi level is
at 0 eV (intrinsic case) in Fig. 1(b), a series of Dirac-type bands with
linear dispersion form the nodal-line structure in the ($k_{x}$, $k_{y}$%
)-plane around $k_{z}$=0, which is similar to the bands in ZrSiS\cite{22,
23, 24}. As $k_{z}$ approaching $\pi $, however, along Z-R line, a new Fermi
surface emerges as the band across the Fermi energy as shown in Fig. 1 (b).

Figure 2 (a) and (b) display representative $\rho _{yx}$-$H$ and $\rho _{xx}$%
-$H$ curves at selected temperatures. Detailed information is given in SMs\cite{51}.
As seen in Fig. 2(a), the $\rho _{yx}$($H$) curves exhibit three remarkable
features. First, the nonlinearity for $\rho _{yx}$($H$)\ develops as the
temperature decreases at low magnetic field. This feature indicates the
carriers from multiband are involved in the transport as temperature
decreases\cite{12,14}. Second, the $\rho _{yx}$($H$) curves have positive
slope for $T$$>$ 40 K, which indicates the majority of carries are $p$-type
at high temperature. Third, for $T$$<$ 40 K, the initial positive slope of $%
\rho _{yx}$($H$) curves tends to change sign as $H$ increases from 0 T to 9
T, which indicates the compensated semimetal behavior with balance of $n$%
-type and $p$-type carriers at low temperature\cite{12,14}. In Fig. S2,
temperature-dependent Hall coefficient, $R_{H}(T)$ = $d\rho _{yx}(T)$/$dH $
at zero field limit shows a hump structure at $T$ $\sim $106 K, which
indicates the minimum value of the total carrier density as temperature
crosses 106 K\cite{51}. To further verifying the deductions, we analyze the Hall
conductivity $\sigma _{xy}$=$\rho _{yx}$/($\rho _{yx}^{2}$+$\rho _{xx}^{2}$)
and use the two-component model to extract the intrinsic carrier densities
and carrier mobilities at various temperatures\cite{9, 24}.
\begin{equation}
\sigma _{xy}=(\frac{n_{1}\mu _{1}^{2}}{1+u_{1}^{2}B^{2}}+\frac{n_{2}\mu
_{2}^{2}}{1+u_{2}^{2}B^{2}})eB.  \label{hall-1}
\end{equation}

Here, $n_{1}$($n_{2})$, $\mu _{1}$($\mu _{2})$ denote the carrier density
[negative(positive)for electrons(hole)] and relevant in-plane mobility, 
respectively. In Fig. 2(c) and (d),
the open symbols and the solid lines label the experiment data and the
fitting results, respectively. The model fits well for the experimental data
at all fixed temperatures. Figure 2 (e) and (f) show the temperature-dependent
fitting parameters. At the lowest temperature $T$= 5 K, the magnitude of $%
n_{1}$ and $n_{2}$ is nearly balanced. As temperature increases, $n_{1}$\
and $\mu _{1}$\ decrease smoothly. In contrast, $\mu _{2}$\ shows two sharp
jumps at $T$=80 K and 106 K labeled by the left dashed red and right dashed
green lines in Fig. 2 (f), respectively. $n_{2}$\ sharply decreases and
becomes positive above $T$=106 K with negligible value in comparison with $%
n_{1}$.

\begin{figure}[tbp]
\begin{flushleft}
\includegraphics[width=1\columnwidth]{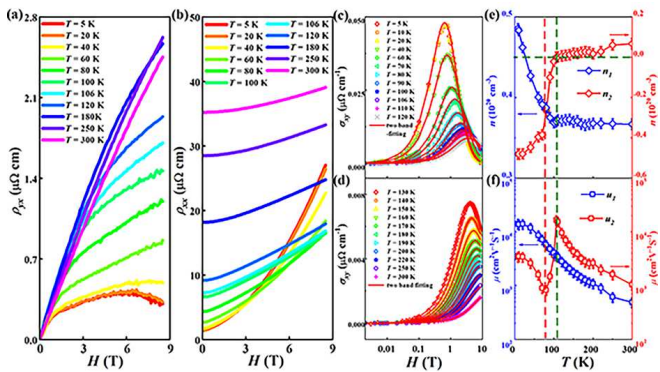}
\caption{(a)-(b) The field dependence of the $\rho_{yx}$($H$) and $\rho_{xx}$($H$) for several temperatures ranging from $T$= 5 K to 300 K. (c)-(d) The field dependence of the Hall conductivity $\sigma_{xy}$($H$) for temperatures ranging from $T$= 5 K to 300 K. Open symbols represent the experimental results and the red solid lines represent the fitting results based on two-component model. (e)-(f) The temperature dependence of fitting parameter, namely, the densities and mobility of the carriers extracted from the two-component model analysis of $\sigma_{xy}$. The left dashed red and right dashed green lines in (f) label two anomalies of the mobility $\mu _{2}$ at 80 K and 106 K respectively.}
\end{flushleft}
\end{figure}

Now, we elaborate the temperature-induced Lifshitz transition can give a
natural interpretation to these transport anomalies. The Hall analyses in
Fig. 2 clarify that the compensated electrons and holes at low temperature
and the majority of $p$-type carriers at high temperature. In combined with
Fig. 1, these behaviors imply that ZrSiSe should have zero Fermi level $E_{F}
$=0 eV at low temperature and the Fermi level should descend at high
temperature. Such Fermi level shift driven by temperature can be induced by
the change of the lattice constants \cite{10} supported by our powder X-ray
diffraction (XRD) measurements (See Fig. S6 and Fig. S7 in SMs for details)\cite{51},
the localization or delocalization induced by the unavoided impurities and
the thermal phonons\cite{49}, and the band renormalization\cite{50} induced
by electron-phonon coupling via self-energy effect. From Fig. 1 (b) and (c),
$p$-type ($n_{1}$\ , $\mu _{1}$) component can be identified to the magenta
Fermi pockets from the red band while the $n$-type ($n_{2}$\ , $\mu _{2}$)
component corresponds to the cyan Fermi pockets from the blue band at $T$%
=5 K. As temperature increase, the first jump of $\mu _{2}$\ emerges at $T$%
=80 K. This can be understood by the band characteristics below and above
80 K. Below 80 K, such as 5 K as shown in Fig. 1(c), the $n$-type cyan Fermi
pockets include both low-mobility carriers with quadratic dispersion (Z-R
line in Fig. 1(b)) and high-mobility carriers with linear dispersion (Z-D
line in Fig. 1(b)). Above 80 K, such as 102 K as shown in Fig. 1(d), the $n$%
-type cyan Fermi pockets only include high-mobility carriers with linear
dispersion (Z-D line in Fig. 1(b)). It indicates that 80 K must correspond to
the Fermi level crossing the quadratic band bottom along the Z-R line in
Fig.1 (b). Meanwhile, $n$-type cyan Fermi pocket splits from one to eight,
which indicates a Lifshitz transition happens at 80 K. As temperature further
increases, the second jump of $\mu _{2}$\ emerges at $T$=106 K. $\mu _{2}$\
reaches a maximal value from 102 K to 106 K. This process corresponds to
shifting the Fermi level from the position labeled by the dotted red line to
the Dirac point labeled by the long dashed green line in Fig. 1(b), because
the carriers at Dirac point have the largest mobility. Above 106 K, the Fermi
level descends below the Dirac point. The $n$-type Fermi pockets disappear
(the blue band is not occupied in Fig. 1(b)), and only the $p$-type Fermi
pockets survive (only the red band is occupied in Fig. 1(b)). It means that
the $n$-type component ($n_{2}$\ , $\mu _{2}$) undergoes the second Lifshitz
transition at 106 K. Note that it seems that two $p$-type components still
exist from Fig. 2(e) and (f), though only one band survives. In the whole
temperature regime, two-component model is adopted to fit the experimental
results. It is natural for each component is not zero. However, above 106 K,
the weight of one component should be very small. It is explicit that $n_{2}$%
\ is insignificantly small in comparison to $n_{1}$. Meanwhile, the $\mu _{1}
$\ and $\mu _{2}$\ have the same trend. Thus, above 106 K, we argue that the $%
n_{1}$\ and $n_{2}$\ may be from the same band (red band in Fig. 1 (b)), and
the small value of $n_{2}$\ may be from the anisotropy of the magenta Fermi
pockets in Fig. 1(e).

STM/S can help to understand the picture extracted from and or beyond the
transport measurements\cite{25, 26, 27, 28, 29, 30}. Figure 3 (b) shows the
topography of a Se cleaved surface and the corresponding fast Fourier
transform (FFT) image at $T$= 4.5 K. Figure 3 (c) and (d) show the zoom-in
view of the topography at $T$= 4.5 K and $T$= 77 K, in which the two
interleaved lattices (Zr and Se) and the Si square net can be well
identified. We next presented in the Fig. 3 (e) and Fig. S9 the line-cut
averaged $dI/dV$ spectra (proportional to the density of states (DOS))
measured at $T$= 4.5 K (black curve) and $T$= 77 K (red curve) along
different scanning paths. The spectra in Fig.3 (e) and Fig. S9 are averaged
by taking several spectra along a Zr-Se-Zr line, and Se-Si-Se line,
respectively. The remarkable difference between two spectrum appears at near
zero bias. A broad V-shape dip emerges at $T$= 4.5 K and disappears at $T$=
77 K. To estimate the temperature-driven change of carriers density, we
present the line-cut map crossing four unit cells (UCs) along Zr-Se-Zr line
at $T$= 4.5 K in Fig. 3 (f). The original $dI/dV$ spectra were shown in Fig.
S8 in SMs\cite{51}. The position-dependent STS spectrum\cite{31} and angle-resolved
photoemission spectroscopy\cite{15, 21} have proven that Zr and Se in ZrSiSe
dominate the holes channel and electron channel, respectively, which is
similar with some semiconductor materials, such as GaAs\cite{32}. Thus, the
imbalance between electron extraction at negative bias and electron
injection at positive bias, \textit{i.e.}, the ratio of $e$-$h$ asymmetric $%
R(r)$, could be directly obtained through the following equation\cite{28,
29, 30}, 
\begin{equation}
R(r)=\frac{\int_{E_{F}}^{-E_{1}}N(r,E)\,dE}{\int_{E_{2}}^{E_{F}}N(r,E)\,dE}.
\label{rr}
\end{equation}%
By integrating the filled and empty states in the vicinity of the $E_{F}$
from -100 mV to 100 mV, the result was shown in Fig. 3g. The $e/h$ ratio
varies between 0.88 and 1.02 within a unit cell, while the overall ratio
takes the value $e/h$=0.95, indicating the nearly perfect compensated band
structure. At $T$= 77 K, as shown in Fig. 3(h) and (i), the orbital texture
changes obviously comparing with the case at $T$= 4.5 K. The overall $e/h$
ratio equals 0.7. Note that the choice of integration limits [-100 meV,
100 meV] guarantees that evolution of the $n$-type band is smooth and
undergoes a Lifshifz transition from the Fermi levels of 4.5 K and 77 K to
-100 meV, respectively. In comparison with the calculated results shown in
Fig. 1 (b), (c), (d) and the two-component model fitting results shown in
Fig. 2 (e), the temperature-induced carriers density evolution obtained from
STM/S measurement follows the same tendency and supports the temperature
induced Lifshifz transitions. The banlanced $e/h$\ ratio also provides a
straightforward explanation of the nonsaturated magnetoresistance (MR) at $T$%
=5 K (See Fig. S10 for details)\cite{33, 34, 51}.
\begin{figure}[tbp]
\begin{flushleft}
\includegraphics[width=1\columnwidth]{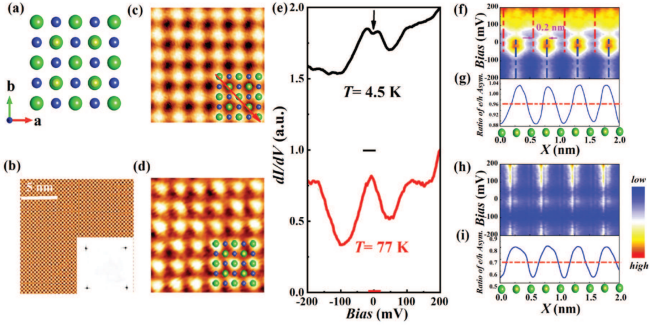}
\caption{(a)Top view of crystal structure of ZrSiSe. (b) Atomically resolved STM topographic image of the cleaved ZrSiSe surface at $T$= 4.5 K with $V_b$ = 600 mV and $I_t$= 200 pA. The inset shows the fast Fourier transform (FFT) image. (c)-(d) Zoom-in views of the cleaved surface at $T$= 4.5 K ($V_b$= 600 mV and $I_t$= 500 pA) and $T$= 77 K ($V_b$= 600 mV and $I_t$= 200 pA). (e) Line-cut averaged $dI/dV$ spectrum acquired in ZrSiSe at $T$= 4.5 K (black curve) and $T$= 77 K (red curve) along the Zr-Se-Zr direction. All spectra are normalized at $V$=200 mV. The black-solid arrow labels the V-shape dip structure. The red-solid arrow in Fig.3(c) labels the line-cut direction. (f) and (h) The STS line-cut map along the Zr-Se-Zr direction at $T$= 4.5 K and 77 K, respectively. (g) and (i) The calculated ratio of $e/h$ asymmetry based on the spectra measurement in (f) and (h), respectively.}
\end{flushleft}
\end{figure}

Now, we turn to understand the V-shape dip of $dI/dV$ at temperature $T$=
4.5 K. Besides SOC, there exists several possible mechanism to induce the
V-shape dip. Since our sample is very clean with the mobility as high as $%
10^{4}cm^{2}V^{-1}S^{-1}$, which excludes the possible disorder effect, such
as the well-known zero-bias anomaly\cite{35, 36, 37, 38, 39, 40, 41, 42}.
The correlated interaction can induce some long-range order, such as
superconducting and excitonic orders. The superconducting order can be
easily excluded, because the V-shape dip structure is robust against the
strong magnetic field in our measurement. Recall that a quite strong Coulomb
interaction can open an excitonic gap to the Dirac nodes in two-dimensional
graphene and the mass enhancement has been recently observed in ZrSiS\cite%
{43, 44, 45, 46}. It is naturally to think that the Dirac nodal-line
structure can be gapped by the similar excitonic order, which results in the
V-shape dip. To verify such a deduction and figure out the relation between
SOC, excitonic instability and the V-shape dip structure. We construct a
simplified model to do a simulation. Let's reconsider the Hall analyses in
Fig. 2 and the electronic structure in Fig. 1 (b) and (c). The minimal model
should includes two parts at least. One part captures the Dirac nodal-line
bands, while another part describes the trivial quadratic band. Note that
the effect of surface bands can merge into the trivial quadratic band,
because surface bands only contribute a large trivial electron pockets
centered at $M$\ point and four small hole pockets around $X$\ point in
surface Brillouin zone\cite{47, 48}. Furthermore, both bulk DOS and surface
DOS are calculated (See Fig. S3 in SMs for details)\cite{51}. Without considering the
excitonic instability, none of them can produce the V-shape dip feature.
\begin{figure}[tbp]
\begin{flushleft}
\includegraphics[width=1\columnwidth]{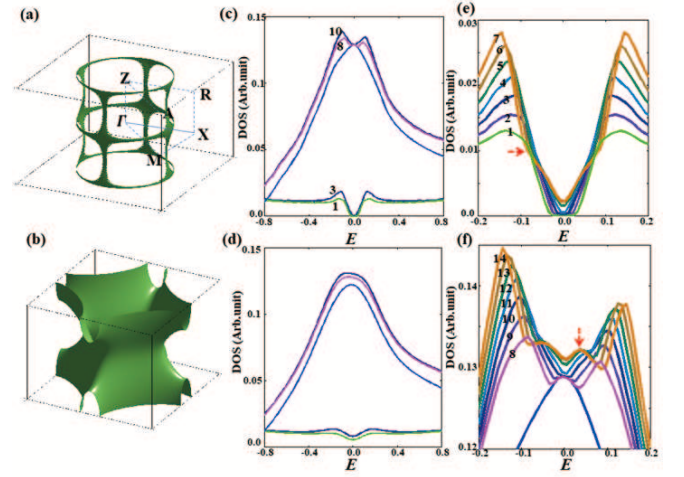}
\caption{(a) and (b)The calculated Dirac nodal-line structure and Fermi surface of trivial quadratic band with the model in SMs, respectively. (c) and (d) The calculated DOS with the model in SMs for $k_B$$T$=0.004 ($T$= 4.5 K) in (c) and $k_B$$T$=0.05 ($T$=77 K) in (d). Bottom curve 1 and 3 are from Dirac nodal-line bands in (a) with ($\lambda_{soc}$, $\Delta_{ex}$)=(0.06, 0) and (0.06, 0.02), respectively. Top curve 8 and 10 are from the sum of blue curve and bottom curves 1 and 3, respectively. (e) and (f) The zoom-in of the bottom and top parts of (c), respectively. Curves 1 and 3 in (e) and (f) are same to these in (c). The curves 2, 4-7 with ($\lambda_{soc}$, $\Delta_{ex}$)=(0.06, 0.01), (0.06, 0.03-0.06). The curves 9, 11-14 have the same values as the curves 2, 4-7. In (a)-(f), other parameters are present in SMs.}
\end{flushleft}
\end{figure}

The detailed theoretical modeling is presented in the SMs\cite{51}, and we only list
the results here. Figure 4(a) and (b) simulates Dirac nodal-line structure and
Fermi surface from the trivial quadratic band. Figure 4 (c) gives several
calculated DOS curves with different SOC $\lambda _{soc}$ and excitonic
order parameter $\Delta _{ex}$. The curve 1 indicates nonzero $\lambda
_{soc} $ can open a gap to the Dirac nodal lines. The curve 3 indicates the
coexistence of both $\lambda _{soc}$ and $\Delta _{ex}$ can suppress the
nodal-line gap opened by $\lambda _{soc}$. The blue color curve gives the
DOS of the trivial quadratic band from Fig. 4 (b). The curve 8 is the sum
from the curve 1 and the blue color one, which is not similar to the 4.5 K
experimental result shown in Fig. 3 (e). It indicates that the sole SOC $%
\lambda _{soc}$ can not explain the experimental result. However, by taking
into account both SOC $\lambda _{soc}$ and the excitonic order parameter $%
\Delta _{ex}$, the cure 10 give similar result in Fig. 3 (e) at 4.5 K.
Furthermore, The thermal broadening effect of $dI/dV$ is proportional to $%
\int d\omega \lbrack -f^{\prime }(\omega -eV)N(\omega )]$ with $f$ and $%
N(\omega )$ the Fermi function and zero-temperature DOS, respectively. The
missing V-shape dip structure of the $dI/dV$ curve at 77 K shown in Fig. 3
(e) is well captured by Fig. 4 (d). To see the effect of both $\lambda
_{soc} $ and $\Delta _{ex}$ clearer, the zoom-in parts of Fig.4 (c) are
shown in Fig. 4 (e) and (f), where more curves with different values of $%
\Delta _{ex}$ are plotted besides the ones in Fig. 4 (c). As $\Delta _{ex}$
increases from zero, the curves 5, 6, 7 show the visible two dip structures
labeled by the red-dashed arrow in Fig. 4 (e). Then, the dip structures of
relevant curves 12, 13, 14 become to deviate from the V-shape labeled by the
red-dashed arrow in Fig. 4 (f). According to our simulation, the optimal
value of $\Delta _{ex}$ to obtain the V-shape dip structure is $\Delta
_{ex}\sim 1/3\lambda _{soc}$, which can strongly suppress the SOC gap
without induce visible two-dip structure.

In conclusion, our transport, STM/S measurements and theoretical analyses
consistently demonstrate that the temperature induces Lifshitz transitions
in ZrSiSe. We also observed a V-shape dip structure around Fermi energy from
the STS spectrum at low temperature. Our theoretical modeling simulation
clarifies the V-shape dip structure is from the co-effect of SOC
and possible excitonic instability. Our observations indicate
correlation interaction in ZrSiSe deserves to attract more attention in the
future studies.

\textbf{Acknowledgements}-This work was supported by the National Key R\&D
Program (2016YFA0300404, 2016YFA0401803, 2017YFA0303201, 2015CB921103,
2019YFA0308602 ), the National Nature Science Foundation of
China (11674326, 11674331, 11774351, 11874357, 11625415, 11374260, U1432139,
U1832141, U1932217), Key Research Program of Frontier Sciences, CAS
(QYZDB-SSW-SLH015), the \textquotedblleft Strategic
Priority Research Program (B)\textquotedblright\ of the Chinese Academy of
Sciences, Grant No. XDB33030100, the `100 Talents Project' of the Chinese Academy of
Sciences, CASHIPS Director's Fund (No. BJPY2019B03) and Science Challenge
Project (No. TZ2016001). A portion of this work was supported by the High
Magnetic Field Laboratory of Anhui Province, the Fundamental Research Funds
for the Central Universities in China, the European Research Council under
the European Union's Seventh Framework Program (FP/2007-2013) through ERC
Grant No. 338957 and by NWO via Spinoza Prize, and the Cluster of Excellence
\textquotedblleft The Hamburg Centre for Ultrafast Imaging (CUI)'\ of the
German Science Foundation (DFG).


\begin{thebibliography}{99}
\bibitem{1} I. M. Lifshitz, Sov. Phys. JETP \textbf{11}, 1130-1135 (1960).

\bibitem{2} Y. Okada, M. Serbyn, H. Lin, D. Walkup, W. Zhou, C. Dhital, M.
Neupane, S. Xu, Y. Wang, R. Sankar, F. Chou, A. Bansil, M. Z. Hasan, S. D.
Wilson, L. Fu and V. Madhavan, Science \textbf{341}, 1496-1499 (2013).

\bibitem{3} C. Liu, T. Kondo, R. M. Fernandes, A. Palczewski, E. D. Mun, N.
Ni, A. N. Thaler, A. Bostwick, E. Rotenberg, J. Schmalian, S. L. Bud'ko, P.
C. Canfield and A. Kaminski, Nat. Phys. \textbf{6}, 419-423 (2010).

\bibitem{4} Z. J. Xiang, G. J. Ye, C. Shang, B. Lei, N. Z. Wang, K. S. Yang,
D. Y. Liu, F. B. Meng, X. G. Luo, L. J. Zou, Z. Sun, Y. Zhang and X. H.
Chen, Phys. Rev. Lett. \textbf{115}, 186403 (2015).

\bibitem{5} L. J. Zhang, C. Y. Guo, X. D.Zhu, L. Ma, G. L. Zheng, Y. Q.
Wang, L. Pi, Y. Chen, H. Q. Yuan and M. L. Tian, Phys. Rev. Lett. \textbf{118
}, 206601 (2017).

\bibitem{6} D. Aoki, G. Seyfarth, A. Pourret, A. Gourgout, A. McCollam, J.
A. N. Bruin, Y. Krupko and I. Sheikin, Phys. Rev. Lett. \textbf{116}, 037202
(2016).

\bibitem{7} G. Bastien, A. Gourgout, D. Aoki, A. Pourret, I. Sheikin, G.
Seyfarth, J. Flouquet and G. Knebel, Phys. Rev. Lett. \textbf{117}, 206401
(2016).

\bibitem{8} Y. Wu, N. H. Jo, M. Ochi, L. Huang, D. Mou, S. L. Bud'ko, P. C.
Canfield, N. Trivedi, R. Arita and A. Kaminski, Phys. Rev. Lett. \textbf{115}%
, 166602 (2015).

\bibitem{9} F. C. Chen, H. Y. Lv, X. Luo, W. J. Lu, Q. L. Pei, G. T. Lin, Y.
Y. Han, X. B. Zhu, W. H. Song and Y. P. Sun, Phys. Rev. B \textbf{94},
235154 (2016).

\bibitem{10} Y. Zhang, C. Wang, L. Yu, G. Liu, A. Liang, J. Huang, S. Nie,
X. Sun, Y. Zhang, B. Shen, J. Liu, H. Weng, L. Zhao, G. Chen, X. Jia, C. Hu,
Y. Ding, W. Zhao, Q. Gao, C. Li, S. He, L. Zhao, F. Zhang, S. Zhang, F.
Yang, Z. Wang, Q. Peng, X. Dai, Z. Fang, Z. Xu, C. Chen and X. J. Zhou, Nat.
Commun. \textbf{8}, 15512 (2017).

\bibitem{11} H. Weng, X. Dai and Z. Fang, Phys. Rev. X \textbf{4}, 011002
(2014).

\bibitem{12} K. Akiba, A. Miyake, Y. Akahama, K. Matsubayashi, Y. Uwatoko
and M. Tokunaga, Phys. Rev. B \textbf{95}, 115126 (2017).

\bibitem{13} G. Manzoni, A. Sterzi, A. Crepaldi, M. Diego, F. Cilento, M.
Zacchigna, Ph. Bugnon, H. Berger, A. Magrez, M. Grioni and F. Parmigiani,
Phys. Rev. Lett. \textbf{115}, 207402 (2015).

\bibitem{14} H. Chi, C. Zhang, G. Gu, D. E. Kharzeev, X. Dai and Q. Li, New
J. Phys. \textbf{19}, 015005 (2017).

\bibitem{15} Q. Xu, Z. Song, S. Nie, H. Weng, Z. Fang and X. Dai, Phys. Rev.
B \textbf{92}, 205310 (2015).

\bibitem{16} Y. Wu, L. Wang, E. Mun, D. D. Johnson, D. Mou, L. Huang, Y.
Lee, S. L. Bud'ko, P. C. Canfield and A. Kaminski, Nat. Phys. \textbf{12},
667-671 (2016).

\bibitem{17} G. Bian, T. R. Chang, R. Sankar, S. Xu, Ha. Zheng, T. Neupert,
C. Chiu, S. Huang, G. Chang, I. Belopolski, D. S. Sanchez, M. Neupane, N.
Alidoust, C. Liu, B. Wang, C. Lee, H. Jeng, C. Zhang, Z. Yuan, S. Jia, A.
Bansil, F. Chou, H. Lin and M. Z. Hasan, Nat. Commun. \textbf{7}, 10556
(2016).

\bibitem{18} L. M. Schoop, M. N. Ali, C. Stra\ss er, A. Topp, A. Varykhalov,
D. Marchenko, V. Duppel, S. S. P. Parkin, B. V. Lotsch and C. R. Ast, Nat.
Commun. \textbf{7}, 11696 (2016).

\bibitem{19} M. N. Ali, L. M. Schoop, C. Garg, J. M. Lippmann, E. Lara, B.
Lotsch and S. S. P. Parkin, Sci. Adv. \textbf{2}, e1601742 (2016).

\bibitem{20} J. Hu, Z. Tang, J. Liu, X. Liu, Y. Zhu, D. Graf, K. Myhro, S.
Tran, C. N. Lau, J. Wei and Z. Mao, Phys. Rev. Lett. \textbf{117}, 016602
(2016).

\bibitem{21} M. M. Hosen, K. Dimitri, I. Belopolski, P. Maldonado, R.
Sankar, N. Dhakal, G. Dhakal, T. Cole, P. M. Oppeneer, D. Kaczorowski, F.
Chou, M. Z. Hasan, T. Durakiewicz and M. Neupane, Phys. Rev. B \textbf{95},
161101 (2017).

\bibitem{22} Y. Huh, E. Moon and Y. B. Kim, Phys. Rev. B \textbf{93}, 035138
(2016).

\bibitem{23} J. Liu and L. Balents, Phys. Rev. B \textbf{95}, 075426 (2017).

\bibitem{24} B. Roy, Phys. Rev. B \textbf{96}, 041113 (2017).

\bibitem{51} See Supplemental Material at [url] for details on the sample
preparation, sample characterization, band structure calculations and the
theoretical modeling of the STS spectra,which includes Refs. [19, $26-37$].

\bibitem{52} P. Hohenberg and W. Kohn, Phys. Rev. \textbf{136}, B864 (1964).

\bibitem{53} W. Kohn and L. J. Sham, Phys. Rev. \textbf{140}, A1133 (1965).

\bibitem{54} G. Kresse and J. Hafner, Phys. Rev. B \textbf{49}, 14251 (1994).

\bibitem{55} G. Kresse and J. Furthm\"{u}ller, Comput. Mater. Sci. \textbf{6}, 15 (1996).

\bibitem{56} G. Kresse and J. Furthm\"{u}ller, Phys. Rev. B \textbf{54}, 11169 (1996).

\bibitem{57} John P. Perdew, Kieron Burke, and Matthias Ernzerhof. Phys. Rev. Lett.
 \textbf{77}, 3865 (1996).

\bibitem{58} P. E. Bl\"{o}chl, Phys. Rev. B \textbf{50}, 17953 (1994).

\bibitem{59} G. Kresse and D. Joubert, Phys. Rev. B \textbf{59}, 1758 (1999).

\bibitem{60} C. Wang and T. Hughbanks, Inorg. Chem. \textbf{34}, 5524 (1995).

\bibitem{61} F. F. Tafti, Q. D. Gibson, S. K. Kushwaha, N. Haldolaarachdhige
and R. J. Cava, Nat. Phys. \textbf{12}, 272-277, (2015).

\bibitem{62} F. F. Tafti, Q. D. Gibson, S. K. Kushwaha, J. W. Krizan,
N. Haldolaarachdhige and R. J. Cava, Proc. Natl. Acad. Sci. USA \textbf{113},
E3475-3481, (2016).

\bibitem{63} X. Luo, F. C. Chen, Q. L. Pei, J. J. Gao, J. Yan, W. J. Lu, P. Tong,
Y. Y. Han, W. H. Song and Y. P. Sun, Appl. Phys. Lett. \textbf{110}, 092401, (2017).

\bibitem{49} D. N. McIlroy, S. Moore, D. Zhang, J. Wharton, B. Kempton, R.
Littleton, M. Wilson, T. M. Tritt and C. G. Olson, J. Phys.: Condens.
Matter. \textbf{16}, L359-356 (2004).

\bibitem{50} Y. S. Kushnirenko, A. A. Kordyuk, A. V. Fedorov, E. Haubold, T.
Wolf, B. B\"{u}chner, and S. V. Borisenko Phys. Rev. B \textbf{96},
100504(R) (2017).

\bibitem{25} A. Piriou, N. Jenkins, C. Berthod, I. M. Aprile and $\varnothing$.
Fischer, Nat. Commun. \textbf{2}, 221 (2011).

\bibitem{26} S. Qiao, X. Li, N. Wang, W. Ruan, C. Ye, P. Cai, Z. Hao, H.
Yao, X. Chen, J. Wu, Y. Wang and Z. Liu, Phys. Rev. X \textbf{7}, 041054
(2017).

\bibitem{27} R. Wu, J. Z. Ma, S. M. Nie, L. X. Zhao, X. Huang, J. X. Yin, B.
B. Fu, P. Richard, G. F. Chen, Z. Fang, X. Dai, H. M. Weng, T. Qian, H.
Ding, and S. H. Pan, Phys. Rev. X \textbf{6}, 021017 (2016).

\bibitem{28} Y. Kohsaka, C. Taylor, K. Fujita, A. Schmidt, C. Lupien, T.
Hanaguri, M. Azuma, M. Takano, H. Eisaki, H. Takagi, S. Uchida, J. C. Davis,
Science \textbf{315}, 1380-1385 (2007).

\bibitem{29} M. Randeria, R. Sensarma, N. Trivedi and F. C. Zhang, Phys.
Rev. Lett. \textbf{95}, 137001 (2005).

\bibitem{30} B. Phillabaum, E. W. Carlson and K. A. Dahmen, Nat. Commun.
\textbf{3}, 915 (2012).

\bibitem{31} K. Bu, Y. Fei, W. Zhang, Y. Zheng, J. Wu, F. Chen, X. Luo, Y.
P. Sun, Q. Xu, X. Dai and Yi Yin, Phys. Rev. B \textbf{98}, 115127 (2018).

\bibitem{32} R. M. Feenstra, J. A. Stroscio, J. Tersoff and A. P. Fein,
Phys. Rev. Lett. \textbf{58}, 1192 (1987).

\bibitem{33} C. Shekhar, A. K. Nayak, Y. Sun, M. Schmidt, M. Nicklas, I.
Leermakers, U. Zeitler, Y. Skourski, J. Wosnitza, Z. Liu, Y. Chen, W.
Schnelle, H. Borrmann, Y. Grin, C. Felser and B. Yan, Nat. Phys. \textbf{11}%
, 645 (2015).

\bibitem{34} M. N. Ali, J. Xiong, S. Flynn, J. Tao, Q. D. Gibson, L. M.
Schoop, T. Liang, N. Haldolaarachchige, M. Hirschberger, N. P. Ong and R. J.
Cava, Nature \textbf{514}, 205 (2014).

\bibitem{35} F. G. Pikus and A. L. Efros, Phys. Rev. B \textbf{51},
16871-16877 (1995).

\bibitem{36} A. L. Efros, B. Skinner and B. I. Shklovskii, Phys. Rev. B
\textbf{84}, 064204 (2011).

\bibitem{37} A. L. Efros, Phys. Rev. Lett. \textbf{68}, 2208 (1992).

\bibitem{38} J. G. Massey and M. Lee, Phys. Rev. Lett. \textbf{75}, 4226
(1995).

\bibitem{39} B. L. Altshuler, A. G. Aronov and P. A. Lee, Phys. Rev. Lett.
\textbf{44}, 1288 (1980).

\bibitem{40} J. P. Eisenstein, L. N. Pfeiffer and K. N. West, Phys. Rev.
Lett. \textbf{69}, 3804 (1992).

\bibitem{41} Z. Y. Jia, Y. H. Song, X. B. Li, K. Ran, P. Lu, H. J. Zheng, X.
Y. Zhu, Z. Q. Shi, J. Sun, J. S. Wen, D. Y. Xing and S. C. Li, Phys. Rev. B
\textbf{96}, 041108 (2017).

\bibitem{42} Y. H. Song, Z. Y. Jia, D. Zhang, X. Y. Zhu, Z. Q. Shi, H. Wang,
L. Zhu, Q. Q. Yuan, H. Zhang, D. Y. Xing and S. C. Li, Nat. Commun \textbf{9}%
, 4071 (2018).

\bibitem{43} V. N. Kotov, B. Uchoa, V. M. Pereira, F. Guinea and A. H.
Castro Neto, Rev. Mod. Phys. \textbf{84}, 1067 (2012).

\bibitem{44} A. N. Rudenko, E. A. Stepanov, A. I. Lichtenstein and M. I.
Katsnelson Phys. Rev. Lett. \textbf{120}, 216401 (2018).

\bibitem{45} M. M. Scherer, C. Honerkamp, A. N. Rudenko, E. A. Stepanov, A.
I. Lichtenstein and M. I. Katsnelson, Phys. Rev. B \textbf{98}, 241112
(2018).

\bibitem{46} S. Pezzini, M. R. van Delft, L. M. Schoop, B. V. Lotsch, A.
Carrington, M. I. Katsnelson, N. E. Hussey and S. Wiedmann, Nat. Phys.
\textbf{14}, 178 (2018).

\bibitem{47} A. Topp, R. Queiroz, A. Gr\"{u}neis, L. M\"{u}chler, A. W.
Rost, A. Varykhalov, D. Marchenko, M. Krivenkov, F. Rodolakis, J. L.
McChesney, B. V. Lotsch, L. M. Schoop and C. R. Ast, Phys. Rev. X \textbf{7}%
, 041073 (2017).

\bibitem{48} B. B. Fu, C. J. Yi, T. T. Zhang, M. Caputo, J. Z. Ma, X. Gao,
B. Q. Lv, L. Y. Kong, Y. B. Huang, P. Richard, M. Shi, V. N. Strocov, C.
Fang, H. M. Weng, Y. G. Shi, T. Qian and H. Ding, Sci. Adv. \textbf{5},
eaau6459 (2019).
\end{thebibliography}
\end{document}